\documentclass[5p]{elsarticle}
\journal{Journal of Magnetism and Magnetic Materials}

\biboptions{sort&compress}
\usepackage{lineno}
\modulolinenumbers[5]

\usepackage{amsmath,bm}

\usepackage[pdftex]{color}
\usepackage{verbatim}
\pdfoutput=1

\usepackage{booktabs,microtype,afterpage} 

\usepackage[colorlinks,plainpages=false,linkcolor=blue,urlcolor=blue,citecolor=blue,pdfpagemode=UseNone,pdfstartview=FitBH]{hyperref}
\usepackage[charter,greekuppercase=italicized]{mathdesign}
\usepackage{graphicx}    
\graphicspath{{./}{figure/}}
\DeclareGraphicsExtensions{.eps,.png,.pdf,.jpg}

\definecolor{red}{rgb}{0.85,.1,0}
\definecolor{green}{rgb}{0.0,0.8,0.0}
\definecolor{orange}{rgb}{1,0.5,0}

\definecolor{blueToni}{rgb}{0.14, 0.36, 0.53}
\definecolor{redToni}{rgb}{0.69, 0.06, 0.086}
\definecolor{dgreen}{RGB}{0,127,0}

\newcommand{\zfmu}{ZF-$\mu$SR}
\newcommand{\tfmu}{TF-$\mu$SR}
\newcommand{\wtfmu}{wTF-$\mu$SR}
\newcommand{\wtf}{wTF}

\newcommand{\musr}{$\mu$SR}

\newcommand{\CRO}{Ca$_3$Ru$_2$O$_7$}
\newcommand{\CFRO}{Ca$_{3}$(Fe$_{x}$Ru$_{1-x}$)$_{2}$O$_{7}$}

\SetExpansion[context = sloppy,shrink = 60]{encoding = {OT1,T1,TS1} }{} 

\begin{document}
% Use nocite{key} to show those entries not cited related to key. If key is * all entries are shown.
%\nocite{*}
\begin{frontmatter}

\makeatletter\renewcommand{\ps@plain}{%
\def\@evenhead{\hfill\itshape\rightmark}%
\def\@oddhead{\itshape\leftmark\hfill}%
\renewcommand{\@evenfoot}{\hfill\small{--~\thepage~--}\hfill}%
\renewcommand{\@oddfoot}{\hfill\small{--~\thepage~--}\hfill}%
}\makeatother\pagestyle{plain}

%\preprint{\textit{Preprint: \today, \now. For internal use only, do not distribute.}}%\linenumbers

\newcommand{\pbcomment}[1]{{\color{red} #1}}

\title{$\mu$SR investigation of the Fe-doped Ca$_{3}$Ru$_{2}$O$_{7}$ polar metal} 
\author[1]{G.~Lamura}
\address[1]{CNR-SPIN, Corso Perrone 24, I-16152 Genova, Italy}
\author[2]{D.~Das}
\address[2]{Laboratory for Muon-Spin Spectroscopy, Paul Scherrer Institut, CH-5232 Villigen PSI, Switzerland}
\author[3]{T.~Shang}
\address[3]{Key Laboratory of Polar Materials and Devices (MOE), School of Physics and Electronic Science, East China Normal University, Shanghai 200241, China}
\author[4]{J.~Peng}
\address[4]{School of Physics, Southeast University, Nanjing 211189, China}
\author[5]{Y.~Wang}
\address[5]{Department of Physics, The Pennsylvania State University, University Park, PA 16802, USA}
\author[5]{Z.~Q.~Mao}
\author[2,6]{T.~Shiroka\corref{cor1}}
\ead{tshiroka@phys.ethz.ch}
\address[6]{Laboratorium f\"ur Festk\"orperphysik, ETH-H\"onggerberg, CH-8093 Z\"urich, Switzerland}

\cortext[cor1]{Corresponding author}

\date{\today}

\begin{abstract}
Ca$_{3}$Ru$_{2}$O$_{7}$ is a polar metal that belongs to the class of
multiferroic magnetic materials. Here, tiny amounts of Fe doping in the 
Ru sites bring about dramatic changes in the electronic and magnetic 
properties and generate a complex $H$-$T$ phase diagram. To date, not 
much is known about the ground state of such a system in the absence of 
magnetic field. By performing muon-spin spectroscopy ($\mu$SR) measurements 
in 5\% Fe-doped \CRO\ single crystals, we investigate its electronic 
properties at a local level. Transverse-field $\mu$SR results indicate 
a very sharp normal-to-antiferromagnetic transition at 
$T_{\mathrm{N}}=79.7(1)$\,K, with a width of only 1\,K. 
Zero-field $\mu$SR measurements in the magnetically ordered state 
allow us to determine the local fields $B_{i}$ at the muon implantation 
sites. By symmetry, muons stopping close to the RuO$_{2}$ planes detect 
only the weak nuclear dipolar fields, while those stopping next to 
apical oxygens sense magnetic fields as high as 150\,mT. In remarkable 
agreement with the nominal Fe-doping, a $\sim 6$\% minority of the 
these muons feel slightly lower fields, reflecting a local magnetic 
frustration induced by iron ions. Finally, $B_{i}$ shows no significant 
changes across the metal-to-insulator transition, close to 40\,K. 
We ascribe this surprising lack of sensitivity to the presence of 
crystal twinning. 
\end{abstract}


%\maketitle
\end{frontmatter}

%\linenumbers

\section{\label{sec:intro}Introduction}
Multiferroic materials are characterized by two or more coexisting 
ferroic orderings, such as ferroelectricity, ferromagnetism, or 
ferroelasticity, which ideally can be mutually controlled and tuned.
As a general feature, conventional multiferroics are electrically 
insulating. However, in rare cases, magnetic and polar orders
can also coexist in metallic systems (despite the seemingly mutually 
exclusive properties of polar order and metallicity).
Collectively such systems are referred to as ``multiferroic magnetic 
materials'' or ``magnetic polar metals'', of which \CRO\ is one 
the most renowned examples~\cite{Cao97,Cao00,Bao08}.
At high temperatures, \CRO\ behaves as a paramagnetic (PM) metal. At low
temperatures, a long-range antiferromagnetic order, with a N\'eel 
temperature $T_{\mathrm{N}} = 56$\,K, sets in. In this case, Ru magnetic
moments align ferromagnetically along the (easy) $a$-axis, while the
RuO$_2$ bilayers stack antiferromagnetically along the $c$-axis (AFM-$a$).
At $T_{\mathrm{MI}} = 48$\,K, the metal-to-insulator (MI) transition
temperature, the resistivity starts to increase and the Ru moments align
along the $b$ axis (AFM-$b$), while preserving their $c$-axis AFM stacking
(see Fig.~7 in Ref.~\cite{Yoshida2005} and Fig.~\ref{fig:Vmag}b further on). 
At low temperatures, the magnetic moment of Ru, as determined by neutron
diffraction, is about 1.8\,$\mu_\mathrm{B}$/Ru and it points along the
$b$ axis~\cite{Bao08}. 

A partial substitution of Ru with Fe (at only 5\%) is able to induce
several dramatic effects. First, it enhances the PM-to-AFM-$a$
transition temperature up to 86\,K~\cite{Ke14,Zhu18,Zhu17,Lei19}.
Secondly, it lowers the $T_{\mathrm{MI}}$ temperature down to 40\,K,
corresponding to an isostructural transition~\cite{Lei19}. Below 40\,K, 
the coexistence of an incommensurate- (IC) and an AFM-$b$ order
(IC + AFM-$b$) sets in. Succinctly, one can portray the magnetic order 
within the RuO$_2$ layers as the spatial repetition of a single unit, 
consisting of two antiparallel AFM domains (AFM-$b$) and two magnetic 
solitons (AFM solitons)~\cite{Lei19}, so as to produce a zero total 
magnetic moment (see Fig.~\ref{fig:Vmag}b).
It was also proposed that, under a zero-magnetic field condition, 
a nearly degenerate spin configuration can occur. In this case, one of 
the two neighboring magnetic solitons in each
perovskite layer could reverse its chirality, thus producing a nonzero 
magnetic moment along the $a$ axis, but not along the $b$ axis.
The Zeeman energy, arising from the application of an external field,
removes such degeneracy, which explains why, for $H \parallel a$, 
an incommensurate ferromagnetic spin structure is found. 
These nearly degenerate spin ground states were ascribed to the magnetic
frustration due to the combined action of diluted Fe$^{3+}$ ions and
magnetocrystalline anisotropy~\cite{Lei19}.

\begin{figure}[thb]
	\centering
    \vspace{-2mm}
	\includegraphics[width=0.3\textwidth]{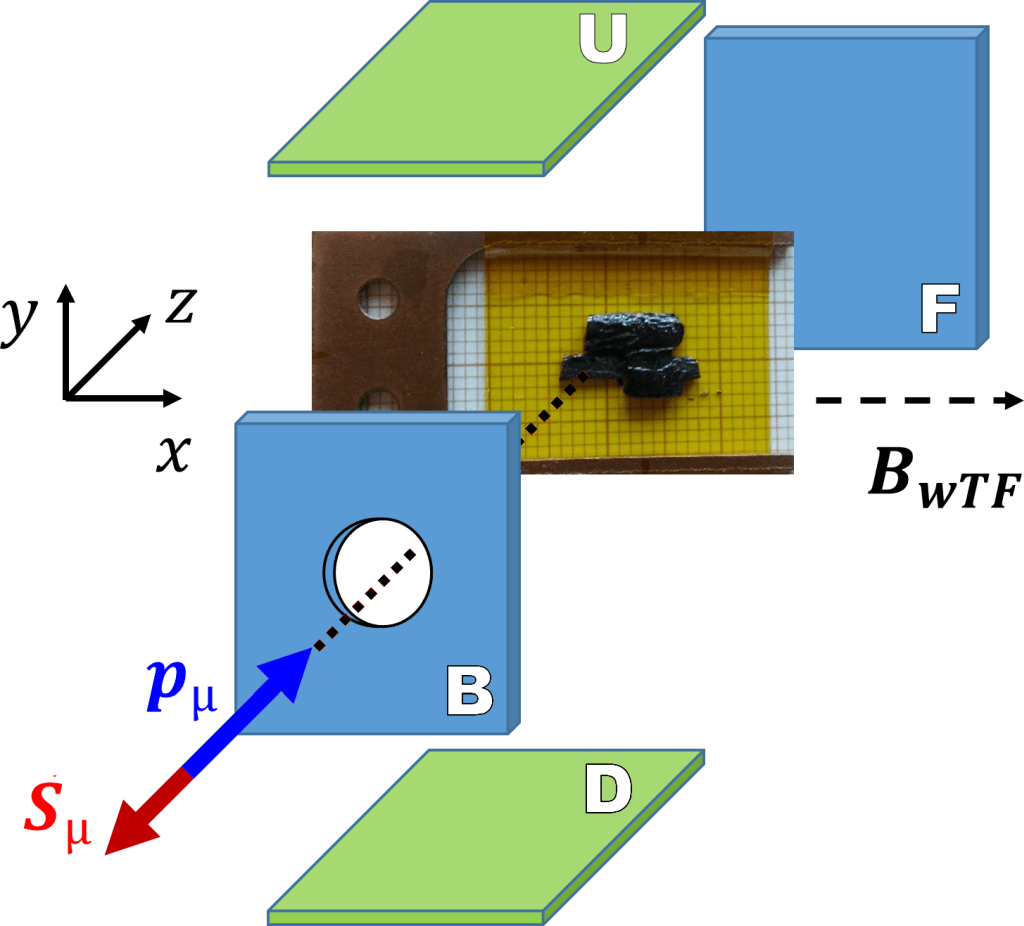} 
	\caption{\label{fig:setup}Geometry of the \zfmu\ experiment, showing
	the detector positions, the sample orientation, as well as the 
	directions of the muon beam and its polarization. The graph paper 
	details the sample dimensions and its align\-ment with respect to 
	the sample-holder axis ($x$-axis in the laboratory frame).}
\end{figure}

In this work, we track down the temperature evolution of the electronic
ground state of Fe-doped \CRO\ under zero-applied field conditions 
(thus leaving the above mentioned degeneracy untouched).  
To achieve this, we rely on the unique capability of zero-field 
muon-spin spectroscopy, which can probe the local magnetic 
field $B_{i}$ at the muon implantation sites, even in the absence of
an applied magnetic field. In this study, we show that, below $T_{\mathrm{N}}$, 
$B_{i}$ lies in the $a\!b$-planes. However, the measured $B_{i}$ shows no differences
across the $T_{\mathrm{MI}}$ transition temperature. We ascribe this
lack of sensitivity to the presence of twinned domains, which make it
impossible to distinguish the local field direction.

\section{\label{sec:sample_growth}Sample growth and characterization}
Fe-doped Ca$_3$Ru$_2$O$_7$ single crystals were grown using the 
floating-zone (FZ) method in a commercial infrared image furnace 
(Cannon Machinery SC2-MDH). The detailed procedures of FZ single-crystal 
growth of ruthenate materials were previously reported in Ref.~\cite{Mao2000}.
The 5\% Fe-doped Ca$_3$Ru$_2$O$_7$ crystals used in this study were  
characterized via x-ray diffraction and confirmed to have the desired 
bilayer ruthenate phase. These crystals and those previously used for 
magnetization-, heat-capacity and neutron-scattering measurements 
(see Ref.~\cite{Ke14}) were taken from the same batch. 
Prior experiments %(Ref.~\cite{Ke14}) 
show the crystals from this batch to have a N\'{e}el temperature 
$T_\mathrm{N} \sim 80$\,K, which coincides with that detected by the 
$\mu$SR measurements reported below. This indicates that the crystals 
used in the current study have a similar quality to those employed in 
previous experiments~\cite{Ke14}.

\begin{figure}[tbh]
	\centering
	\vspace{-2mm}
	\includegraphics[width=0.42\textwidth]{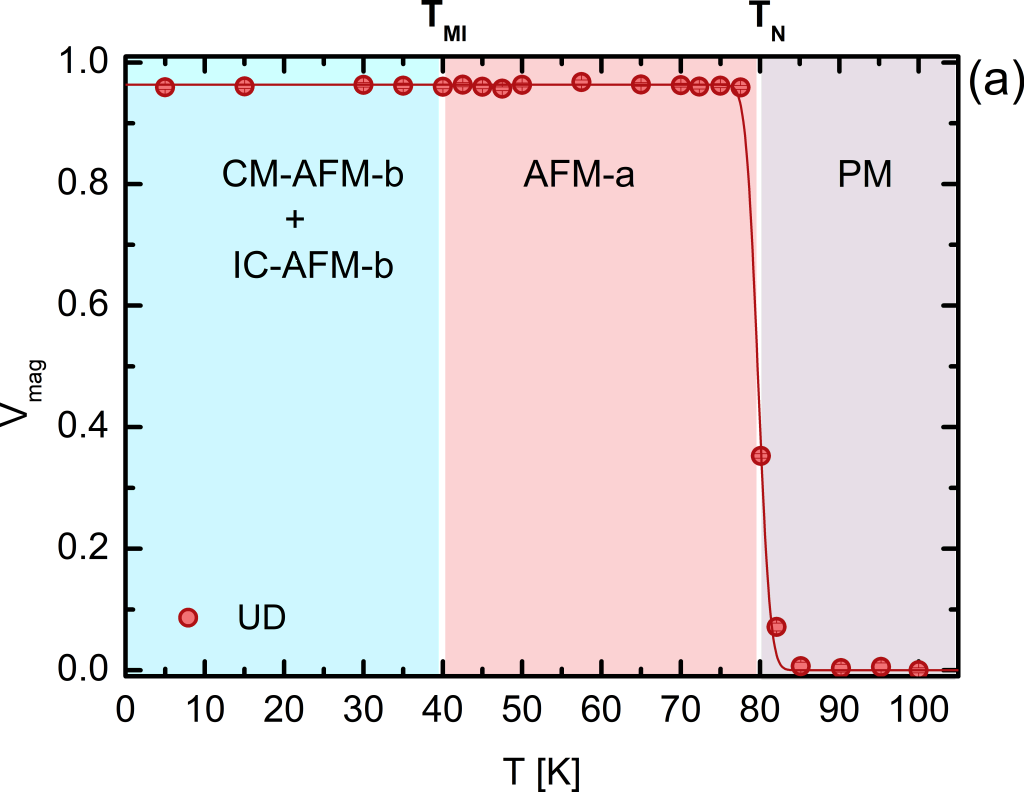}%
	\vspace{2mm}
	\includegraphics[width=0.45\textwidth]{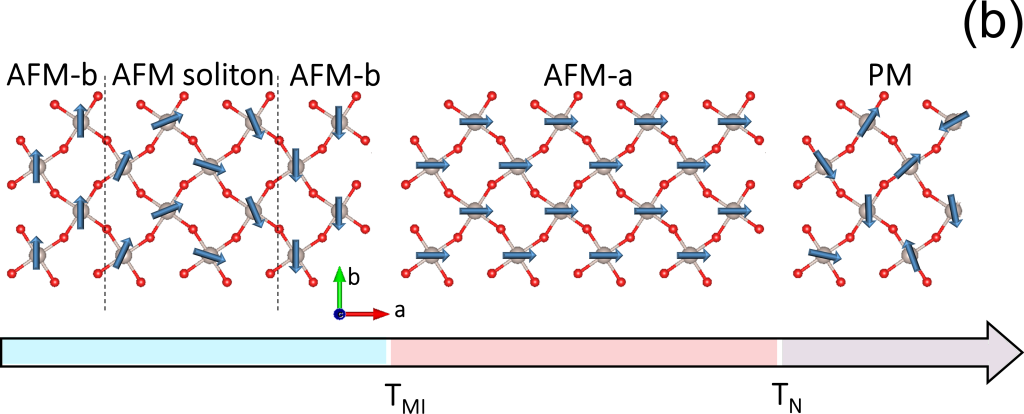}
	\caption{\label{fig:Vmag}(a) Temperature dependence of the magnetic 
	volume fraction, as extracted from the \wtf\ UD asymmetry data. 
	The continuous line represents a numerical fit by means of 
	an \textit{erf}-function, which gives a magnetic transition 
	temperature $T_{\mathrm{N}}=79.7(1)$\,K and a transition width 
	$\Delta T_{\mathrm{N}}=1.2(1)$\,K. (b) In-plane view of the magnetic 
	structures for $T<T_{\mathrm{MI}}$ (commensurate AFM-$b$ coexisting 
	with incommensurate AFM-$b$), $T_{\mathrm{MI}}<T<T_{\mathrm{N}}$ 
	(commensurate AFM-$a$) and $T>T_{\mathrm{N}}$ (PM). In both panels,  
	the colored areas indicate the different magnetic phases of \CFRO\ in zero applied field.}
\end{figure}

\section{\label{sec:results}\musr\ spectroscopy results}
The \musr\ technique uses spin-polarized positive muons implanted
homogeneously over the whole sample volume. Upon implantation,
the incident muons loose their kinetic energy, yet preserve the initial
spin direction (polarization). Therefore, once stopped, they start to 
precess coherently around the local magnetic field at the implantation 
site. The decay positrons, emitted preferentially along the muon-spin 
direction, carry the physical information in the form of precession 
frequency and relaxation rate. Here, we perform two kinds of 
experiments: (i) weak transverse-field- ({w\tfmu}), and (ii) 
ze\-ro\--field ({\zfmu}) muon spectroscopy measurements. The former 
was used to determine the temperature evolution of the magnetic 
volume fraction, while the latter allowed us to study the temperature 
evolution of the magnetically ordered phase.

\subsection{\label{sec:samples}Experimental setup}

The \musr\ experiments were carried out at the general-purpose 
spectrometer (GPS) at the $\pi$M3 beam line of Paul Scherrer 
Institut (PSI), Villigen, Switzerland. A non-rotated muon-spin 
configuration was adopted during all the experiments. The investigated 
sample consisted of a mosaic of homogeneous single crystals positioned 
on a thin Kapton (polyimide) tape, with their $c$ axes parallel to the 
muon momentum $\boldsymbol{p_{\mu}}$. 
The shorter sample sides were oriented parallel 
to the vertical direction defined by the axis of the up--down (UD) 
detectors. The experimental setup, including the detectors and the 
sample orientation, is shown in Fig.~\ref{fig:setup}. 
The relatively large sample thickness ($\simeq 0.5$\,mm) and the use of 
veto counting (which considers only the muons effectively stopped 
in the sample) contributed to an improved signal-to-noise ratio.

\subsection{\label{sec:vmag} Magnetic volume fraction determined via \wtfmu}
The temperature evolution of the magnetic volume fraction was determined 
by means of a {w\tfmu} experiment, as detailed in \ref{app:A}.
In the PM-phase, all the implanted muons precess at the 
Larmor frequency. Once the temperature decreases below the magnetic 
transition $T_\mathrm{N}$, only those muons implanted in the gradually
shrinking PM-phase still precess at the frequency of the applied field. 
In this case, the temperature evolution of the magnetic volume fraction 
$V_{\mathrm{mag}}(T)$ can be derived from the equation: 
\begin{equation}
	\label{eq:wTeq}
	V_{\mathrm{mag}}(T) = 1-\frac{a_{\mathrm{PM}}(T)}{a_{\mathrm{PM}}(T_{\mathrm{max}})},
\end{equation}
where $T_{\mathrm{max}}$ is the maximum temperature attained during 
the experiment, here located in the PM state ($T_{\mathrm{N}} < T_{\mathrm{max}}$), 
while the ratio $\frac{a_{\mathrm{PM}}(T)}{a_{\mathrm{PM}}(T_{\mathrm{max}})}$ 
represents the paramagnetic fraction, here determined from the muon-spin asymmetry. 
The temperature-dependent magnetic volume fraction, as measured by the 
UD detectors, is shown in Fig.~\ref{fig:Vmag}a. It is worth noting 
that: (i) the magnetic transition is very sharp. This is confirmed 
by a fit of the experimental data with a Gaussian distribution of local 
transition temperatures around the average value $T_{\mathrm{N}}$:
\begin{equation}
\label{eq:vmag}
V_{\mathrm{mag}}(T) =\frac{V_{\mathrm{mag}}(0)}{2} \, \left[ 1 - \mathrm{erf} \left( \frac{T-T_{\mathrm{N}}}{\sqrt 2 \Delta T_{\mathrm{N}}} \right) \right].
\end{equation} 
Here, erf is the standard error function \cite{erf}, 
$\Delta T_{\mathrm{N}}$ is the %root mean square deviation 
width of such distribution, and $V_{\mathrm{mag}}(0)$ is the 
zero-temperature limit of the magnetic volume fraction. The best fit 
gives $T_{\mathrm{N}}=79.7(1)$\,K, with an average transition width 
$\Delta T_{\mathrm{N}}=1.2(1)$\,K. (ii) Below $T_{\mathrm{N}}$, the 
volume fraction of the ordered phase is almost unity, 
$V_{\mathrm{mag}}=0.964(4)$, indicative of a very high sample quality.

\subsection{\label{sec:vmag-prop}\zfmu: Properties of the magnetically ordered phase}

\subsubsection{\label{sec:Bdirection}Internal field direction in the magnetically ordered phase}
\begin{figure}[tbh]
	\centering
   %\vspace{-2mm}
	\includegraphics[width=0.42\textwidth]{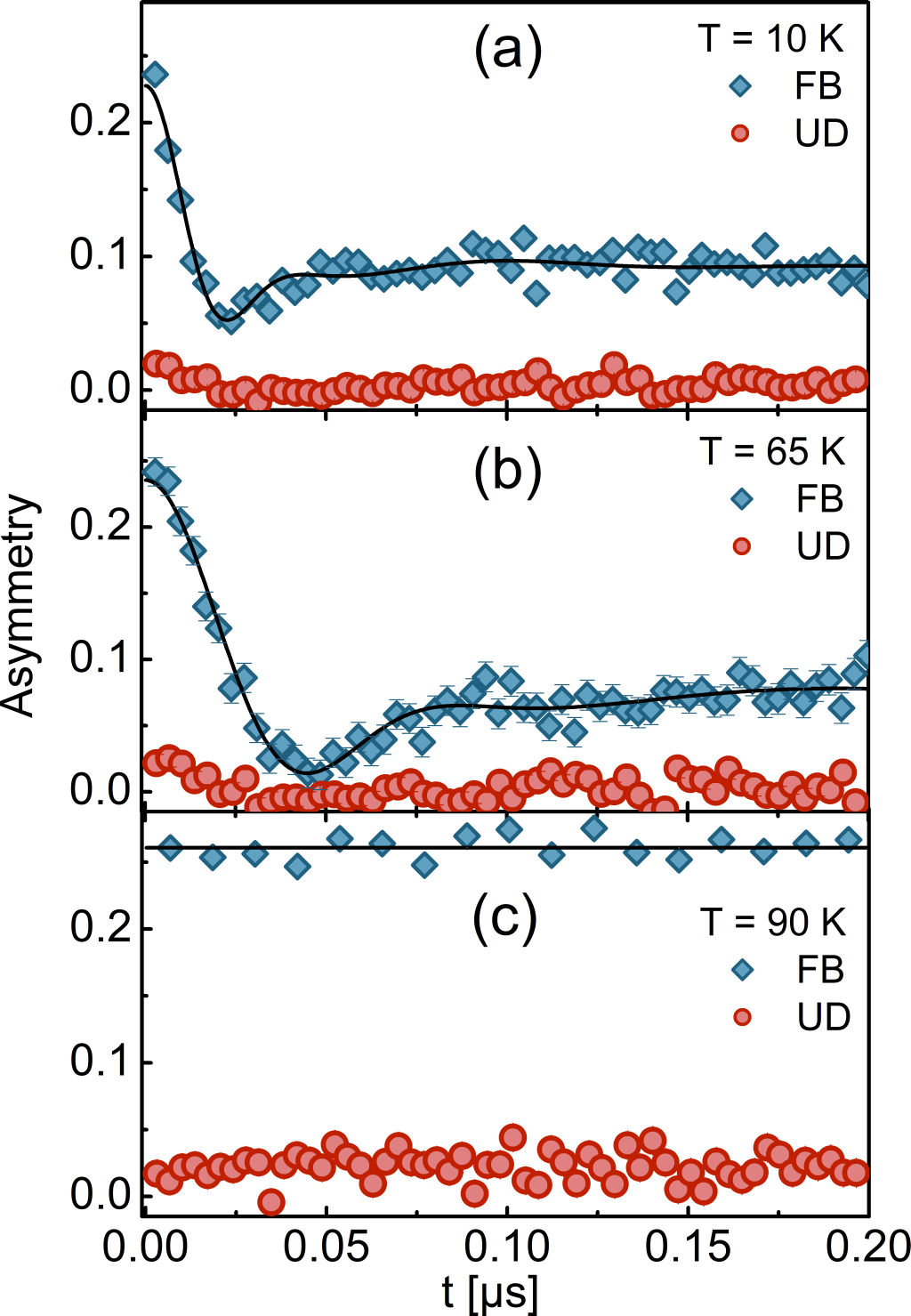}
	\caption{\label{fig:AsymLt}Short-time \zfmu\ asymmetry, as measured 
	by the FB and UD detector pairs, at selected temperatures in a 
	non-rotated muon-spin configuration. Note the constant zero signal 
	in the UD detectors.}
\end{figure}

Figure~\ref{fig:AsymLt} shows the ZF-$\mu$SR results obtained in a 
non-rotated muon-spin configuration at three representative temperatures:  
10\,K, 65\,K, and 90\,K, i.e., below $T_{\mathrm{MI}}$, 
at intermediate $T_{\mathrm{MI}}<T<T_{\mathrm{N}}$, and well above $T_{\mathrm{N}}$. 
In this configuration, by inspecting the asymmetry data as recorded in
the forward-backward (FB) and up--down (UD) detector pairs, the direction
of the internal magnetic field at the muon implantation sites can be
inferred (see, e.g., Ref.~\cite{Tran2018}). The depicted datasets show 
two distinct features. The UD detectors exhibit an almost zero 
asymmetry~\cite{UDasym}, independent of temperature. The FB detectors, 
instead, show a nonzero temperature- and time-dependent asymmetry. 

The absence of an UD asymmetry indicates that the internal magnetic 
field $B_{i}$, as probed by the implanted muons, cannot be parallel 
to the crystal $c$-axis. Indeed, this would have implied a zero- and 
a constant (nonzero) asymmetry in the UD and FB detector pairs,
respectively. Since this is not the case, we conclude that the direction
of the internal magnetic field at the muon implantation sites lies
necessarily in the laboratory $x\!y$ plane, here corresponding to the
sample's $a\!b$ plane (see Fig.~\ref{fig:setup}).
Unfortunately, the Ca$_{3}$Ru$_{2}$O$_{7}$ lattice parameters $a$
($\simeq 537$\,pm) and $b$ ($\simeq 522$\,pm) are very close. This,
and the additional presence of crystal twinning, makes the in-plane
crystal orientation hard to establish, thus ruling out the possibility 
to determine the exact internal field direction within the $a\!b$-planes.

\begin{figure}[bht]
	\centering
   %\vspace{-2mm}
	\includegraphics[width=0.42\textwidth]{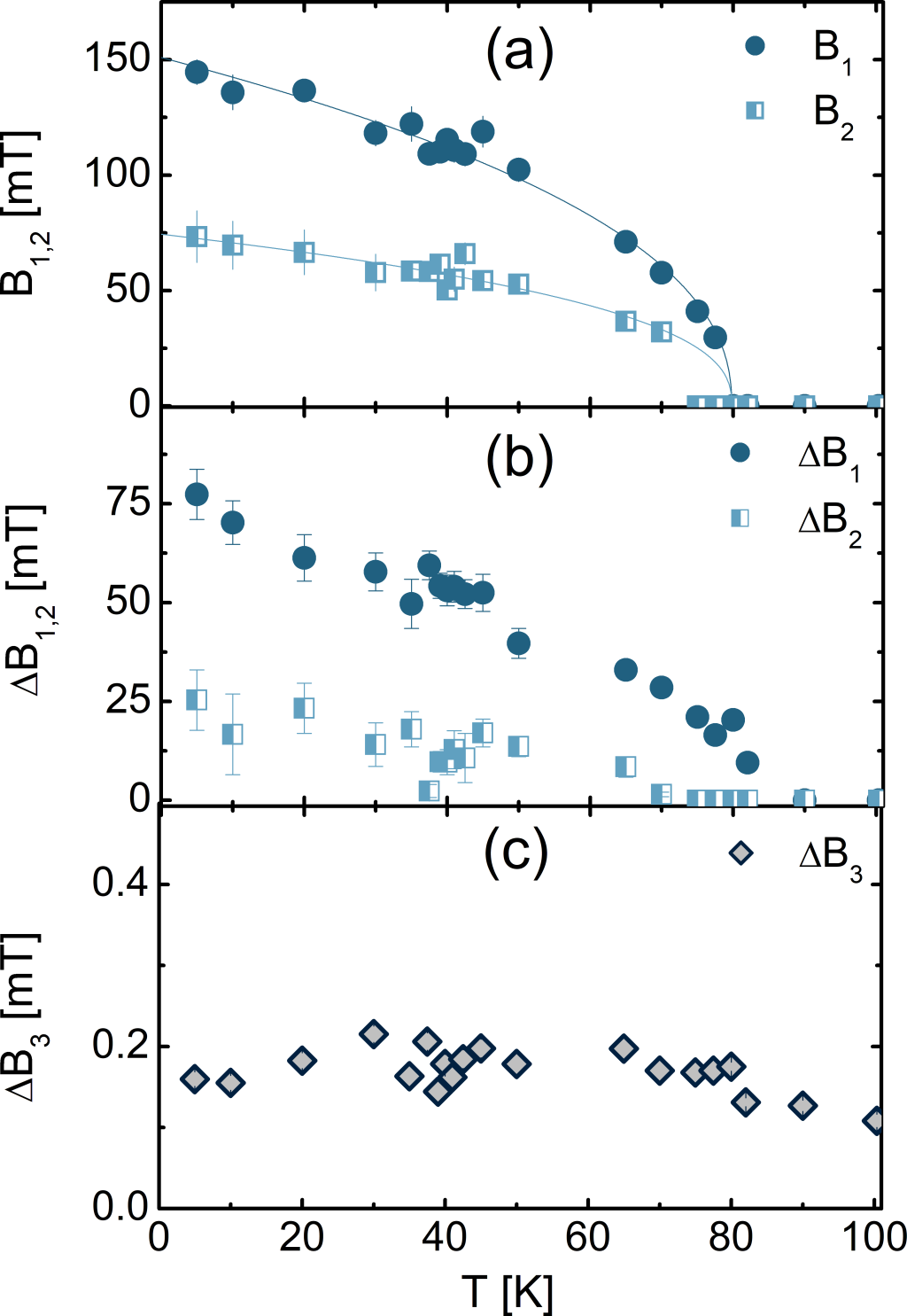}
	\caption{\label{fig:bmu}\zfmu\ parameters as resulting from fits of 
	the time-dependent asymmetry. Top and middle panels show the local 
	magnetic fields $B_{i}$ and field widths $\Delta B_{i}$. The 
	continuous lines in the top panel are representative of the 
	standard mean-field theory. Bottom panel shows the longitudinal 
	relaxation rate.}
\end{figure}

\subsubsection{Temperature dependence of the staggered magnetization}
The temperature dependence of the magnetic order parameter can be
extracted from the FB asymmetry, with typical datasets being shown
in Fig.~\ref{fig:AsymLt}. From these measurements one can distinguish
two different temperature regimes: 
(a) a low-temperature regime ($T \leq T_{\mathrm{N}}$), characterized
by highly damped oscillations (indicative of a magnetically ordered
state) superimposed on a slow-decaying relaxation, observable only
at long times; and (b) a high-temperature regime ($T > T_{\mathrm{N}}$),
where only a slow-decaying relaxation is observed. To track these 
changes across the whole temperature range, the asymmetry data were 
fitted to the following function~\cite{chi2}:
\begin{equation}\label{eq:pt}
  \begin{split}
    A(t) = a_{1}\cos \left(\gamma_{\mu} B_{1} + \varphi \right) \cdot e^{-\sigma^{2}_{1}t^2/2} +\\
    + a_{2}\cos \left(\gamma_{\mu} B_{2}+ \varphi \right) \cdot e^{-\sigma^{2}_{2}t^2/2} +\\ 
    + a_{3}\cdot e^{-\sigma^{2}_{3}t^2/2}.
  \end{split}
\end{equation}
Here, the transverse (oscillating) component is represented by two
cosine functions, with $a_{1,2}$, $B_{1,2}$ and $\sigma_{1,2}$ being
the initial amplitude, local field, and damping rates, respectively. 
%\tcr{The need for two functions is demanded by the Fourier transform of 
%the time-domain signal, which shows the presence of two broad peaks at 
%70 and 136\.mT.}
The longitudinal component was fitted to a relaxing Gaussian term, with
$a_{3}$ and $\sigma_{3}$ being the initial amplitude and the relaxation
rate, respectively. This fit model suggests some important observations. 
There are three nonequivalent muon implantation sites (1, 2, and 3),
with relative weights of 54(4), 6(1), and 40(1)\%, respectively, 
mostly independent of temperature.
Muons implanted in 1 and 2 probe a nonzero local field that, at 5\,K, is
as high as 150(5) and 73(11)\,mT, respectively. The highly damped
oscillations at very short times, could be consistently fitted
only by means of a cosine function (a fit with a zeroth-order Bessel 
function proved unsatisfactory). Muons implanted in the third site 
experience only the nuclear dipolar fields, since their long-time 
asymmetry data can only be fitted with a Gaussian-like decay, as 
detailed in \ref{app:B}. 

Figure~\ref{fig:bmu} shows the temperature dependence of the fitted
parameters. The top panel reports the behavior of the internal magnetic
field values, $B_{1,2}$, the middle panel the widths 
of the corresponding field distribution, $\Delta B_{1,2}$, and the
bottom panel the second moment of the field distribution sensed by the
muons implanted in the third site, $\Delta B_{3}$. We recall that, 
for local fields with a Gaussian distribution, the second moment is 
given by $\Delta B_{i}=\sigma_{i} / \gamma_{\mu}$, with $i = 1$, 2, 3.
It is worth noting several interesting features: (i) below $T_\mathrm{N}$, both $B_{1,2}$
and $\Delta B_{1,2}$ increase with decreasing temperature. (ii) the data
are well fitted by a standard mean-field model. (iii) Interestingly,
no significant changes could be detected across $T_{\mathrm{MI}}$.
(iv) $\Delta B_{3}$ is about 0.15\,mT, irrespective of the
temperature, as expected for nuclear dipolar-field contributions. 
We discuss the possible origin of the three local fields in 
the next section.

\section{\label{sec:discussion} Discussion}
\CRO\ has a perovskite structure, similar to that of cup\-ra\-tes and
manganites. By using similar arguments to those put forward in Ref.~\cite{Coneri10}, 
we expect two main muon implantation sites. As observed in all perovskites, 
a first site corresponds to muons which bind themselves to the apical
oxygen of the Ru-O tetrahedra (eight sites per unit cell). A second
site is expected to lie in the RuO$_2$ planes, at the center of a
rhombus formed by four oxygen atoms (eight sites per unit cell).
Therefore, at least two distinct precession frequencies, with a
relative weight of 50\%, are expected. For symmetry reasons, the field
probed by the muons implanted in the RuO$_2$ planes should be zero. 
As to the first site, by assuming the dipolar contribution to the local 
field as dominant, one can make a rough estimate of the expected local 
field (in this case, we consider copper-oxide compounds 
for a comparison~\cite{Coneri10}). Thus, by considering an ordered Ru 
moment of $1.8$\,$\mu_\mathrm{B}$~\cite{Bao08} and by scaling it to 
the magnetic moment of copper, the field sensed by muons bound to the 
apical oxygens is expected to be in the 90--120\,mT range.

Our experimental results can be interpreted by assuming two distinct 
precession frequencies and a Gaussian relaxing component. Since the 
latter exhibits a temperature-independent behavior (with a relative 
weight of 40\%), it accounts for the muons implanted in the RuO$_2$ planes. 
Here, for symmetry reasons, the electronic dipolar contribution 
due to Ru ions is zero and, therefore, only the $T$-independent 
Ru nuclear dipolar fields dominate. This is confirmed also by 
the \wtf\ measurements, where small applied fields are sufficient to 
quench the tiny nuclear contribution (see \ref{app:A}).
By making an educated guess, the two precessing components of the recorded 
asymmetry can be associated with the muons bound to the apical oxygens.  
A minority of these muons, with a relative weight of 6\%, show a lower 
frequency and most likely correspond to apical oxygens near the Fe-Ru 
substitutions ($\sim$ 5\%). Indeed, in this case, we expect a local 
frustration of the magnetic order, which lowers the local field.
On the other hand, the higher frequency and larger weight of the other 
component is due to muons bound to the apical oxygens, far from Fe 
ions, which sense the original field of the unsubstituted Ru sites. 
This interpretation implies that, because of the Fe-Ru substitution, 
$\sim 5$\% of muon implantation sites are inequivalent to the rest. 
In addition, the electronic dipolar field at sites 1 and 2 lies 
in the $a\!b$ plane, as confirmed by the absence of 
contributions to the UD asymmetry below $T_{\mathrm{N}}$. Unfortunately, 
because of the presence of crystal twinning, we cannot resolve the 
in-plane orientation of such field. This could also be the reason
for the absence of visible changes in $B_{1,2}(T)$ and 
$\Delta B_{1,2}(T)$ across $T_{\mathrm{MI}}$ (at 48\,K).

\section{\label{sec:conclusion}Conclusion}
In summary, we investigated the temperature evolution of the zero-field
electronic ground state of \CFRO\ by means of \musr\ spectroscopy.
First, by applying a weak transverse field, we confirm that this compound
undergoes a very sharp magnetic transition from a paramagnetic metallic
phase to an AFM-$a$ ordered state at 79.7(1)\,K. By means of \zfmu\
spectroscopy we determine the local magnetic fields at two different 
implantation sites, corresponding to muons bound to apical oxygens and 
to those lying close to the RuO$_2$ planes. 
In agreement with a nominal 5\% iron doping, a minority of muons bound 
to apical oxygens exhibit lower precession frequencies, reflecting a local 
frustration of the magnetic order due to Fe-Ru substitutions.
Unfortunately, the presence of crystal twinning hinders the detection 
of spin reordering, known to occur across the MI transition at 48\,K. 
A future study, using uniaxial strain, should allow us to eliminate 
the twin domains and thus track the change in magnetic structure across 
$T_{\mathrm{MI}}$.

\section{\label{sec:acknowledgments}Acknowledgments}
Y.W.\ and Z.Q.M.\ acknowledge support from the Penn State 2D Crystal
Consortium-Materials Innovation Platform (2DCC-MIP) funded by NSF
(cooperative agreement DMR-153\-99\-16). T.Sh.\ was in part supported
by the Swiss National Science Foundation under Grant No.\ 200021-169455.

% -----------------------------------------------------------
\appendix

\section{\label{app:A}\wtfmu\ data analysis}

In a standard \wtfmu\ experiment the magnetic field is applied 
perpendicular to the initial $\mu^{+}$ spin polarization. The term 
``weak'' has two important meanings \cite{wTF2017,wTF2019,Yaouanc2011}:
(i) the applied field $B_\mathrm{app}$ should be small compared to 
the local field $B_{i}$ in the magnetically ordered phase. 
In our case, $B_\mathrm{app}= 3$\,mT is smaller than 
$B_{i} \sim 20$\,mT, the typical local field normally 
observed in AFM-ordered compounds~\cite{Yaouanc2011}. (ii) When the 
applied field is orthogonal also to the $\mu^{+}$ momentum (as in the 
present case), the field intensity must be sufficiently low to cause 
only a negligible displacement of the incoming muon beam. High transverse 
fields would deviate the muon beam too much, which, consequently, would 
miss the sample. In our case, the chosen value of $B_\mathrm{app}$ 
implies a beam shift of less than 1\,mm.

\begin{figure}[tbh]
	\centering
    \vspace{-2mm}
	\includegraphics[width=0.42\textwidth]{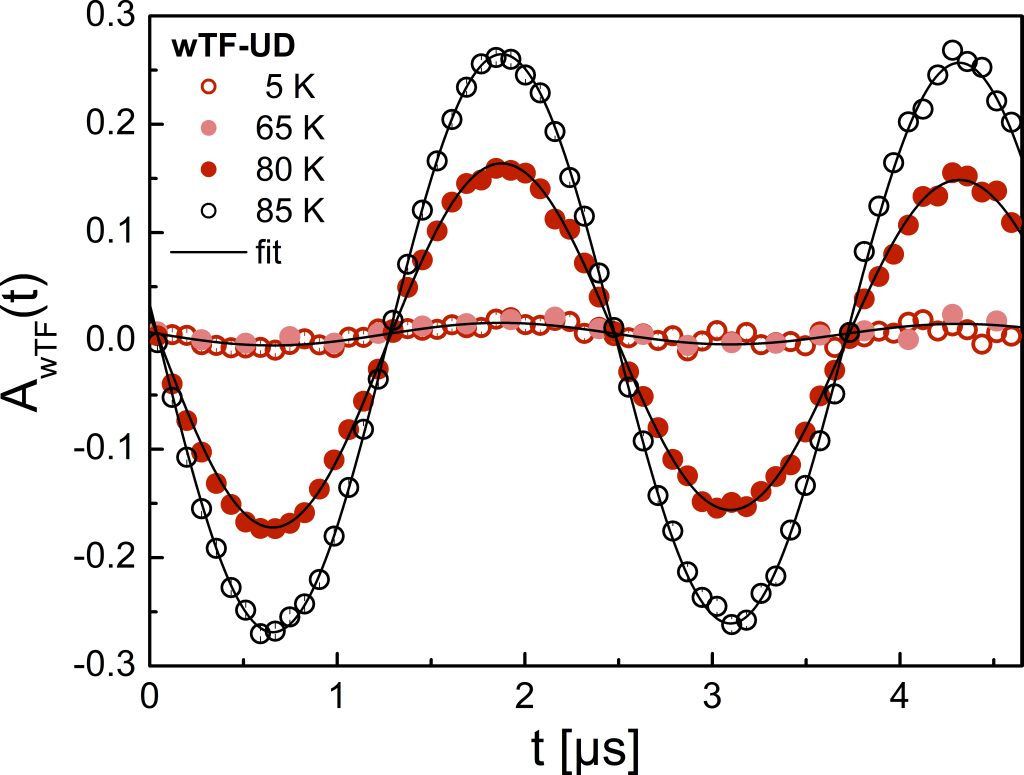}
	\caption{\label{fig:wTFAsym}\wtfmu\ asymmetry as measured 
	in the UD detectors at selected temperatures in a 
	non-rotated muon-spin configuration.}
\end{figure}

Several \wtf-$\mu$SR datasets were collected  
between 5 and 100\,K in a rotated spin configuration. In a transverse 
configuration, in principle, one can use equivalently the FB or UB 
time-dependent asymmetries to determine the evolution of the PM 
fraction with temperature. 
Since for $T<T_{\mathrm{N}}$ the local field is parallel to the 
$a\!b$-planes (see par.~\ref{sec:Bdirection}), this 
implies a null signal in the UD pair. Hence, the asymmetry data from 
such detectors are particularly meaningful for following the evolution 
of the PM fraction. Figure~\ref{fig:wTFAsym} shows the \wtf\ UD 
asymmetry at selected temperatures. The data were analyzed by means 
of the function: 
\begin{equation}
	\label{eq:wTeq}
	A_{\mathrm{wTF}}(t) = a_{\mathrm{PM}} \cos(\gamma_{\mathrm{\mu}} 
	B_{\mathrm{\mu}} t + \phi) \cdot e^{-\lambda_{} t} + a_{\mathrm{tail}} \cdot e^{-\lambda_\mathrm{tail} t}.
\end{equation}
Here, $A_{\mathrm{wTF}}(t)$ is the time-dependent asymmetry, 
$a_{\mathrm{PM}}$ is the amplitude of the oscillating paramagnetic
component, $\lambda_{\mathrm{PM}}$ is the corresponding damping rate. 
Contrary to the ZF case (see \ref{app:B}), here the nuclear dipole 
field is quenched by the applied field and, therefore, the relaxation 
is mainly Lorentzian.
$\gamma_{\mathrm{\mu}} B_{\mathrm{\mu}}$ is the angular Larmor frequency 
in the applied field, $\phi$ is a phase offset, and $\gamma_\mu\!=\!2\pi \times 135.5$\,MHz/T 
is the muon gyromagnetic ratio. The relaxing component 
$a_{\mathrm{tail}}(t)$ accounts for a tiny nonzero offset in the AFM 
phase, possibly due to a small sample misalignment~\cite{UDasym}.

\section{\label{app:B}Long-time \zfmu\ asymmetry}
\begin{figure}[htb]
	\centering
	\includegraphics[width=0.42\textwidth]{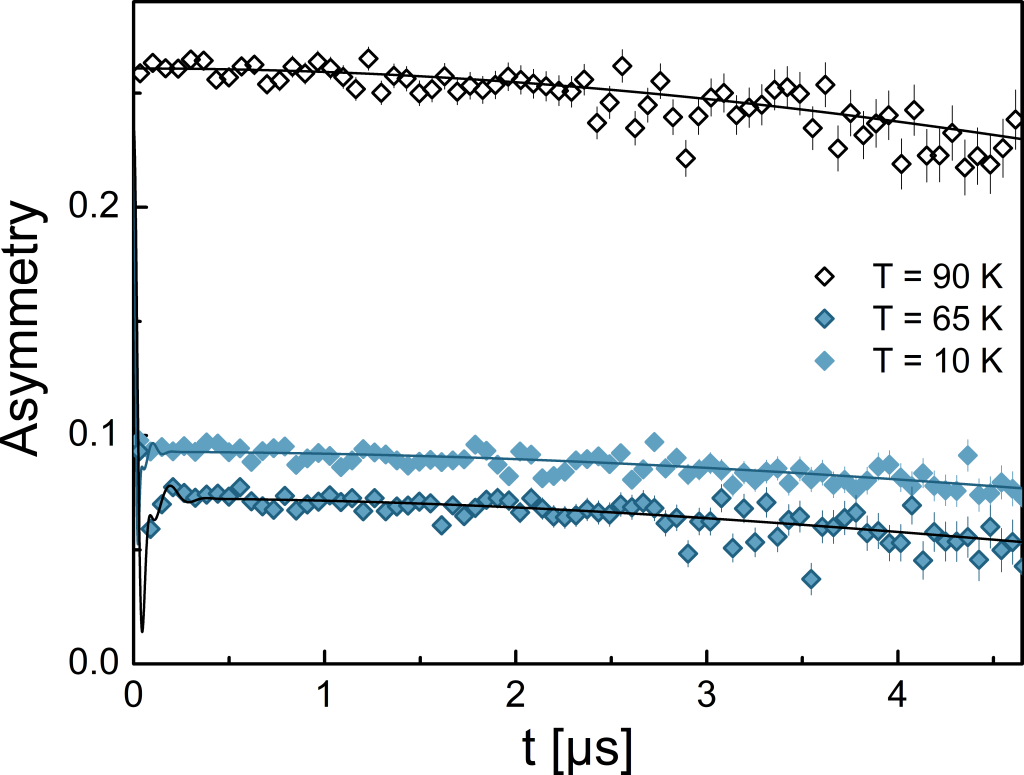} 
	\caption{\label{fig:AsymHt}The long-time \zfmu\ asymmetry at $T = 10$,
	65, and 90\,K, as recorded in the FB detectors, is compatible with a 
	Gaussian relaxation.} 
\end{figure}

Figure~\ref{fig:AsymHt} shows the long-time \zfmu\ asymmetry at
$T = 10$, 65, and 90\,K, respectively, as recorded in the FB detectors.
The asymmetry profile is compatible with a Gaussian-like relaxation,
whose value is essentially independent of temperature, as expected for 
the muon-spin relaxation induced by nuclear dipolar fields.

%

% Uncomment the line below to hide the article titles. TS
%\bibliographystyle{apsrev4-2}
%\bibliography{CaRuO_v3}

\begin{thebibliography}{10}
\expandafter\ifx\csname url\endcsname\relax
  \def\url#1{\texttt{#1}}\fi
\expandafter\ifx\csname urlprefix\endcsname\relax\def\urlprefix{URL }\fi
\expandafter\ifx\csname href\endcsname\relax
  \def\href#1#2{#2} \def\path#1{#1}\fi

\bibitem{Cao97}
G.~Cao, S.~McCall, J.~E. Crow, R.~P. Guertin, Observation of a metallic
  antiferromagnetic phase and metal to nonmetal transition in
  {Ca}$_3${Ru}$_2${O}$_7$, Phys. Rev. Lett. 78 (1997) 1751--1754.
\newblock \href {https://doi.org/10.1103/PhysRevLett.78.1751}
  {\path{doi:10.1103/PhysRevLett.78.1751}}.

\bibitem{Cao00}
G.~Cao, K.~Abboud, S.~McCall, J.~E. Crow, R.~P. Guertin, Spin-charge coupling
  for dilute {La}-doped {Ca}$_3${Ru}$_2${O}$_7$, Phys. Rev. B 62 (2000)
  998--1003.
\newblock \href {https://doi.org/10.1103/PhysRevB.62.998}
  {\path{doi:10.1103/PhysRevB.62.998}}.

\bibitem{Bao08}
W.~Bao, Z.~Q. Mao, Z.~Qu, J.~W. Lynn, Spin valve effect and magnetoresistivity
  in single crystalline {Ca}$_3${Ru}$_2${O}$_7$, Phys. Rev. Lett. 100 (2008)
  247203.
\newblock \href {https://doi.org/10.1103/PhysRevLett.100.247203}
  {\path{doi:10.1103/PhysRevLett.100.247203}}.

\bibitem{Yoshida2005}
Y.~Yoshida, S.-I. Ikeda, H.~Matsuhata, N.~Shirakawa, C.~H. Lee, S.~Katano,
  Crystal and magnetic structure of {Ca}$_3${Ru}$_2${O}$_7$, Phys. Rev. B 72
  (2005) 054412.
\newblock \href {https://doi.org/10.1103/PhysRevB.72.054412}
  {\path{doi:10.1103/PhysRevB.72.054412}}.

\bibitem{Ke14}
X.~Ke, J.~Peng, W.~Tian, T.~Hong, M.~Zhu, Z.~Q. Mao,
  Commensurate-incommensurate magnetic phase transition in the {Fe}-doped
  bilayer ruthenate {Ca}$_3${Ru}$_2${O}$_7$, Phys. Rev. B 89 (2014) 220407.
\newblock \href {https://doi.org/10.1103/PhysRevB.89.220407}
  {\path{doi:10.1103/PhysRevB.89.220407}}.

\bibitem{Zhu18}
M.~Zhu, T.~Hong, J.~Peng, T.~Zou, Z.~Q. Mao, X.~Ke, Field-induced magnetic
  phase transitions and memory effect in bilayer ruthenate
  {Ca}$_3${Ru}$_2${O}$_7$ with {Fe} substitution, J. Phys.: Condens. Matter 30
  (2018) 075802.
\newblock \href {https://doi.org/10.1088/1361-648x/aaa626}
  {\path{doi:10.1088/1361-648x/aaa626}}.

\bibitem{Zhu17}
M.~Zhu, J.~Peng, T.~Hong, K.~Prokes, T.~Zou, Z.~Q. Mao, X.~Ke, Field-induced
  metastability of the modulation wave vector in a magnetic soliton lattice,
  Phys. Rev. B 95 (2017) 134429.
\newblock \href {https://doi.org/10.1103/PhysRevB.95.134429}
  {\path{doi:10.1103/PhysRevB.95.134429}}.

\bibitem{Lei19}
S.~Lei, S.~Chikara, D.~Puggioni, J.~Peng, M.~Zhu, M.~Gu, W.~Zhao, Y.~Wang,
  Y.~Yuan, H.~Akamatsu, M.~H.~W. Chan, X.~Ke, Z.~Mao, J.~M. Rondinelli,
  M.~Jaime, J.~Singleton, F.~Weickert, V.~S. Zapf, V.~Gopalan, Comprehensive
  magnetic phase diagrams of the polar metal
  {Ca}$_3$({Ru}$_{0.95}${Fe}$_{0.05}$)$_{2}${O}$_7$, Phys. Rev. B 99 (2019)
  224411.
\newblock \href {https://doi.org/10.1103/PhysRevB.99.224411}
  {\path{doi:10.1103/PhysRevB.99.224411}}.

\bibitem{Mao2000}
Z.~Mao, Y.~Maeno, H.~Fukazawa, Crystal growth of {Sr}$_2${Ru}{O}$_4$, Mater.
  Res. Bull. 35~(11) (2000) 1813--1824.
\newblock \href {https://doi.org/https://doi.org/10.1016/S0025-5408(00)00378-0}
  {\path{doi:https://doi.org/10.1016/S0025-5408(00)00378-0}}.

\bibitem{erf}
$\mathrm{erf}(x)=\frac{2}{\sqrt \pi } \cdot \int_{0}^{x} e^{-t^2} \mathrm{d}t$.

\bibitem{Tran2018}
L.~M. Tran, M.~Babij, L.~Korosec, T.~Shang, Z.~Bukowski, T.~Shiroka, Magnetic
  phase diagram of {Ca}-substituted {EuFe}$_2${As}$_2$, Phys. Rev. B 98 (2018)
  104412.
\newblock \href {https://doi.org/10.1103/PhysRevB.98.104412}
  {\path{doi:10.1103/PhysRevB.98.104412}}.

\bibitem{UDasym}
The tiny residual asymmetry observed in the UD detectors is most likely due to
  a small misalignment of the muon spin with respect to the $c$-axis direction
  of the sample.

\bibitem{chi2}
The chosen function shows a slightly lower $\chi^2$ with respect to a linear
  combination of two Kubo-Toyabe functions. Most importantly, it matches better
  the experimental data at short times. Other fitting models, such as a simple
  Kubo-Toyabe, or its linear combination with a Gaussian decay, largely miss
  the short-time asymmetry behavior, which demands the use of two oscillating
  functions, as evinced by two broad peaks at 70 and 136\,mT in the Fourier
  transform (not shown).

\bibitem{Coneri10}
F.~Coneri, S.~Sanna, K.~Zheng, J.~Lord, R.~De~Renzi, Magnetic states of lightly
  hole-doped cuprates in the clean limit as seen via zero-field muon spin
  spectroscopy, Phys. Rev. B 81 (2010) 104507.
\newblock \href {https://doi.org/10.1103/PhysRevB.81.104507}
  {\path{doi:10.1103/PhysRevB.81.104507}}.

\bibitem{wTF2017}
R.~Khasanov, Z.~Guguchia, A.~Maisuradze, D.~Andreica, M.~Elender, A.~Raselli,
  Z.~Shermadini, T.~Goko, F.~Knecht, E.~Morenzoni, A.~Amato, High pressure
  research using muons at the {P}aul {S}cherrer {I}nstitute, High Press. Res.
  36~(2) (2016) 140--166.
\newblock \href {https://doi.org/10.1080/08957959.2016.1173690}
  {\path{doi:10.1080/08957959.2016.1173690}}.

\bibitem{wTF2019}
O.~K. Forslund, D.~Andreica, Y.~Sassa, H.~Nozaki, I.~Umegaki, E.~Nocerino,
  V.~Jonsson, O.~Tjernberg, Z.~Guguchia, Z.~Shermadini, R.~Khasanov, M.~Isobe,
  H.~Takagi, Y.~Ueda, J.~Sugiyama, M.~M{\aa}nsson, Magnetic phase diagram of
  {K$_{2}$Cr$_{8}$O$_{16}$} clarified by high-pressure muon spin spectroscopy,
  Sci. Rep. 9 (2019) 1141.
\newblock \href {https://doi.org/10.1038/s41598-018-37844-5}
  {\path{doi:10.1038/s41598-018-37844-5}}.

\bibitem{Yaouanc2011}
A.~Yaouanc, P.~{Dalmas de R\'eotier}, Muon Spin Rotation, Relaxation, and
  Resonance: Applications to Condensed Matter, Oxford University Press, Oxford,
  2011.

\end{thebibliography}

\end{document}